\documentclass[aps,prl,twocolumn,showpacs,preprintnumbers,amsmath,amssymb]{revtex4-1}

 \def\frac#1#2{{#1\over #2}}
 \def\l{{\lambda}}

 \newcommand{\Tr}{\text{Tr}}

\def\be{\begin{equation}}
\def\ee{\end{equation}}
\newcommand{\bea}{\begin{eqnarray}}
\newcommand{\eea}{\end{eqnarray}}
\newcommand{\ket}[1]{\left| #1 \right>}
\newcommand{\bra}[1]{\left< #1 \right|}
\newcommand{\ketbra}[1]{\left|#1\right>\left<#1\right|}

\usepackage{color}
\usepackage{graphicx}
\usepackage{dcolumn}
\usepackage{bm}
\usepackage{hyperref}

\begin{document}

\title{Krylov complexity of modular Hamiltonian evolution}

\author{Pawel Caputa$^1$, Javier M. Magan$^2$, Dimitrios Patramanis$^1$ and Erik Tonni$^3$ }

\affiliation{$^1$Faculty of Physics, University of Warsaw, Pasteura 5, 02-093 Warsaw, Poland }
\affiliation{$^2$Instituto Balseiro, Centro Atomico Bariloche 8400-S.C. de Bariloche, Rio Negro, Argentina}
\affiliation{$^3$SISSA and INFN Sezione di Trieste, via Bonomea 265, 34136, Trieste, Italy}

\date{\today}

\begin{abstract}
We investigate the complexity of states and operators evolved with the modular Hamiltonian by using the Krylov basis. 
In the first part, we formulate the problem for states and analyse different examples, including quantum mechanics,  two-dimensional conformal field theories and random modular Hamiltonians, focusing on relations with the entanglement spectrum. We find that the modular Lanczos spectrum provides a different approach to quantum entanglement, opening new avenues in many-body systems and holography. In the second part, we focus on the modular evolution of operators and states excited by local operators in two-dimensional conformal field theories. We find that, at late modular time, the spread complexity is universally governed by the modular Lyapunov exponent $\lambda^{mod}_L=2\pi$ and is proportional to the local temperature of the modular Hamiltonian. Our analysis provides explicit examples where entanglement entropy is indeed not enough, however the entanglement spectrum is, and encodes the same information as complexity.

\end{abstract}

\maketitle

\section*{Introduction and summary}
In recent years quantum complexity has become a new exciting area within quantum many-body systems, quantum gravity and quantum field theory, see e.g. \cite{Aaronson,Chen:2021lnq,Chapman:2021jbh}. It provides a new perspective on the structure of quantum states as well as quantum dynamics, complementing that of quantum information. It is also instrumental in understanding black holes in holography \cite{Ma} and quantum gravity \cite{Harlow:2013tf,Susskind,Stanford:2014jda,Brown:2015bva}. In this last context, it was argued by Susskind that ``entanglement is not enough'' \cite{Susskind:2014moa}, in particular if one seeks to understand aspects of the long time regime of chaotic systems and black holes. Complexity measures were then proposed as fine-grained probes of dynamics at late times. For this reason, most research in this direction has focused on properties of a real time, Hamiltonian evolution of complexity measures, trying to test their supremacy to entanglement measures. On the other hand, it is known that entanglement entropy contains only a small fraction of information about bipartite entangled states, and that the entanglement spectrum is a much more fine-grained measure of their structure \cite{LiHaldane:2008}. In this vein, it is natural to ask for more direct relations between complexity and entanglement, and better characterize what types of entanglement measures are enough for the study of holography and quantum black holes. This is one of our main motivations in this work.

To this end, we focus on arguably the most promising (in relation to the problems mentioned above) definition of complexity called, Krylov \cite{Parker:2018yvk} or spread complexity \cite{Balasubramanian:2022tpr}, that can be applied both to quantum operators and states. This measure was inspired by pioneering works on operator size, quantum chaos and thermalisation in many-body systems \cite{Roberts:2018mnp,Qi:2018bje}, and it has produced a burst of activity and interest in recent years \cite{Barbon:2019wsy,Magan:2020iac,Rabinovici:2020ryf,Dymarsky:2021bjq,Kar:2021nbm,Caputa:2021sib,Patramanis:2021lkx,Caputa:2021ori,Caputa:2022eye,Caputa:2022yju,Khetrapal:2022dzy,Kundu:2023hbk,Bhattacharjee:2022vlt,Hornedal:2022pkc,Takahashi:2023nkt,Hornedal:2023xpa,Carabba:2022itd,Avdoshkin:2022xuw,Rabinovici:2022beu,Lin:2022rbf,Balasubramanian:2022dnj,Balasubramanian:2022gmo,Camargo:2022rnt,Hashimoto:2023swv,Camargo:2023eev,Bhattacharyya:2023dhp,Patramanis:2023cwz,Lv:2023jbv,Muck:2022xfc,Adhikari:2022whf,Kim:2021okd,Erdmenger:2023shk,Rabinovici:2023yex,Nandy:2023brt,Bhattacharjee:2022qjw,Gautam:2023pny,Dixit:2023fke,Hashimoto:2023swv,Patramanis:2023cwz,Iizuka:2023pov,Bhattacharyya:2023dhp}. As shown in \cite{Balasubramanian:2022tpr}, Krylov or spread complexity defines complexity as the minimal amount of spread of the wavefunction in the Hilbert space. Such minimization is universally accomplished for a finite amount of time by the so-called Krylov basis, that arises via the Lanczos recursion method \cite{LanczosVish} (to be reviewed below). Some highlights of  Krylov complexity in many body systems are the demonstration of the exponential growth with the universal Lyapunov exponent, together with the idea that Krylov complexity bounds the growth of out-of-time-ordered correlators \cite{Parker:2018yvk,Gu:2021xaj}, the derivation of the linear growth regime of complexity \cite{Barbon:2019wsy}, the geometric approach and connection to generalized coherent states \cite{Caputa:2021sib}, the ability to codify quantum chaos and fine-grained properties of the spectrum \cite{Balasubramanian:2022tpr,Balasubramanian:2022dnj,Erdmenger:2023shk}, and the recent holographic demonstrations that it can reproduce the volume of black hole interiors, a.k.a. the volumes of Einstein-Rosen bridges, see \cite{Lin:2022rbf,Erdmenger:2023shk,Rabinovici:2023yex} for the JT gravity cases, and \cite{Balasubramanian:2022gmo} for the case of general relativity in general dimensions.
 
Given these recent developments, and with the aim of exploring the relation between entanglement and complexity, in this work we expand the Krylov or Lanczos approach in two new directions. First, generalising previous work on the time evolution of the thermofield double state (TFD) \cite{Balasubramanian:2022tpr}, we define and study the complexity of modular evolution in generic bipartite entangled states. We show that this measure is controlled by the entanglement spectrum of the reduced density matrix. Equivalently it is controlled by the modular Lanczos spectrum, which interestingly contains the very same information, and might be taken as a new characterization of the entanglement structure. Indeed the first modular Lanczos coefficient is the entanglement entropy itself, while the second is the (square) of capacity of entanglement \cite{Yao:2010woi,DeBoer:2018kvc}. We will then analyze this quantity in several examples. For random states, the modular Hamiltonian is random, and we discuss how entanglement entropy is codified in the plateau of the modular complexity evolution, and how modular complexity is also sensitive to the Page curve \cite{Page:1993df}. In the second part we will discuss  modular growth and evolution of quantum operators. In holography this paves a way towards a precise measure of complexity of bulk reconstruction. Exploiting again the power of generalised coherent states as well as modular two-point correlators in two-dimensional (2d) CFTs, we will derive a universal growth of spread complexity of modular evolution characterised by Lyapunov exponent $\lambda^{mod}=2\pi$ and the scrambling time governed by an effective local temperature of the modular Hamiltonian for a single interval as well as two intervals in free fermion CFT.

Overall, this approach makes it clear why ``entanglement is not enough'' \cite{Susskind:2014moa}, while at the same time it also suggests that a slight but insightful modification may solve the issue at stake, and that indeed entanglement spectrum is enough.
\section*{Spread Complexity}
For completeness, we begin with a brief review of the spread complexity  \cite{Balasubramanian:2022tpr}. The starting point of the discussion is the unitary evolution of an initial quantum state $\ket{\psi_0}$ with time-independent Hamiltonian $H$
\be
\ket{\Psi(t)}=e^{-iHt}\ket{\psi_0}.\label{InState}
\ee
Generically, this evolution spreads the state $\ket{\psi_0}$ in the Hilbert space of the model, making it more complex. While the amount of the spread depends on the choice of basis, we can quantify the complexity of this process by minimizing the spread of the wavefunction over all choices of basis. The result of this minimization, at least for a finite period of time, brings us to the so-called Krylov basis. This basis, denoted below by $\ket{K_n}$, is obtained via the Gram-Schmidt orthogonalisation procedure on the subspace of all the powers of $H$ applied to $\ket{\psi_0}$. The iterative procedure to achieve this is called the Lanczos algorithm \cite{LanczosVish} and it can be written as
\be
\ket{A_{n+1}}=(H-a_n)\ket{K_n}-b_n\ket{K_{n-1}},
\ee 
where $\ket{K_n}=b^{-1}_n\ket{A_n}$, $b_0=0$ and the first vector coincides with our initial state $\ket{K_0}=\ket{\psi_0}$. The key role in this story is played by Lanczos coefficients $a_n$ and $b_n$ that control the  dynamics and are defined as
\be
a_n=\bra{K_n}H\ket{K_n},\qquad b_n=\langle A_n|A_n\rangle^{1/2}.\label{LanczCoeffDeff}
\ee
The algorithm stops as soon as any of the $b_n=0$, which signifies that no more independent basis vectors can be constructed. After running this iterative algorithm, we can expand the state in the Krylov basis
\be\label{expand}
\ket{\Psi(t)}=\sum_n \psi_n(t)\ket{K_n}.
\ee
By construction, the coefficients of this expansion satisfy a discrete Schrodinger equation
\be
i\partial_t\psi_n(t)=a_n\psi_n(t)+b_n\psi_{n-1}(t)+b_{n+1}\psi_{n+1}(t),\label{SEq}
\ee
that also highlights the fact that the Hamiltonian is tridiagonal in the Krylov basis, with tridiagonal elements given by the Lanczos coefficients. Finally, if we are able to solve this equation, the spread complexity is computed as the average value of $n$ in the probability distribution $p_n(t)\equiv|\psi_n(t)|^2$, namely
\be
\mathcal{C}(t)=\sum_{n}n\,p_n(t).\label{SpreadC}
\ee
Clearly, solving \eqref{SEq} is the main step and it requires the knowledge of the Lanczos coefficients. They are in fact encoded in the return amplitude (the Loschmidt amplitude)
\be
S(t)=\langle \Psi(t)|\Psi(0)\rangle=\bra{\psi_0}e^{iHt}\ket{\psi_0}=\sum_n\mu_n\frac{t^n}{n!}.\label{Moments}
\ee
Its moments $\mu_n=\langle \psi_0|(iH)^n|\psi_0\rangle$ allow us to extract Lanczos coefficients that are related via polynomial equations e.g., the first two are (see more in Appendix A)
\bea
a_0=-i\mu_1,\qquad b^2_1=\mu^2_1-\mu_2.
\label{abmus}
\eea
Inversely, the knowledge of the Lanczos coefficients allows the computation of the moments of the Hamiltonian. Therefore, since the Lanczos coefficients play such a pivotal role, it is important to understand their physical meaning and how different phenomena are encoded in their scaling with $n$.

We conclude this introduction with two remarks. Firstly, an important class of initial states $\ket{\psi_0}$ is given by the TFD state \cite{Takahashi:1996zn}. Denoting by $\ket{n}$ the eigenstate of the Hamiltonian with energy $E_n$ this state reads
\be
\ket{\psi_\beta}=\frac{1}{\sqrt{Z(\beta)}}\sum_n e^{-\frac{\beta}{2} E_n}\ket{n}_L\otimes\ket{n}_R,\label{TFDst}
\ee
where $Z(\beta)$ is the partition function at temperature $T=1/\beta$. The TFD state is the canonical purification of the thermal density matrix $\rho=e^{-\beta H}$. It is then interesting to consider the time evolution of \eqref{TFDst} with the Hamiltonian of a single copy, say $H_L$, especially in the context of black holes \cite{Hartman:2013qma,Papadodimas:2015xma}. For this evolution the return amplitude becomes the analytically continued partition function
\be
S(t)=\frac{Z(\beta-it)}{Z(\beta)},
\ee
whose modulus squared is the spectral form factor, a key object in the field of quantum chaos \cite{GUHR1998189}. This way, the Lanczos coefficients as well as spread complexity are directly probing the spectrum of the evolving Hamiltonian, and  they codify the fine-grained aspects such as spectral rigidity and the universality class of the chaotic model \cite{Balasubramanian:2022tpr,Balasubramanian:2022dnj,Erdmenger:2023shk}. Indeed, for chaotic systems with no degeneracies the Lanczos spectrum of this process contains exactly the same information as the spectrum itself. The main idea of this work is to generalise this TFD example to reduced density matrices and modular Hamiltonian evolution. 

Secondly, the Krylov complexity of the operator growth \cite{Parker:2018yvk} can be studied in a complete analogy with the discussion above. The only non-trivial step is the choice of the inner-product in the space of operators that allows us to map Heisenberg evolution of an operator $\mathcal{O}(t)$ to a state $|\mathcal{O}(t))$. The crucial information about the operator growth is then captured by the return amplitude that corresponds to a two-point correlator $(\mathcal O|\mathcal{O}(t))$. Along these lines, below we will consider operator growth as well as the dynamics of CFT states excited by local operators under modular Hamiltonian evolution. They will involve return amplitudes based on modular two-point functions in 2d CFTs.  
\section*{Modular Spread Complexity}
We now consider spread complexity of modular Hamiltonian evolution. As reviewed above, we start with a pure state $\left|\Psi_0\right>$ in some Hilbert space $\mathcal{H}$. We then pick a sub-system $A$ and its complement $A^c$, and assume a Hilbert space decomposition $\mathcal{H}= \mathcal{H}_A\otimes  \mathcal{H}_{A^c}$, so that we can write $\left|\Psi_0\right>$ in the Schmidt form
\begin{equation}
\left|\Psi_0\right>=\sum_{j}\sqrt{\lambda_j}\left|j\right>_A\left|j\right>_{A^c},\label{SchmDC}
\end{equation}
where $\ket{j}$ are basis vectors in $A$ (and the complement). As usual, we define the reduced matrix $\rho_A$ of the sub-region $A$ as well as the modular Hamiltonian $H_A$ by
\begin{equation}
\rho_A=\text{Tr}_{A^c}\left(\left|\Psi_0\right>\left<\Psi_0\right|\right)\equiv e^{-H_A}.
\end{equation}
The Schmidt coefficients in \eqref{SchmDC} describe the spectrum $\lambda_j$ of $\rho_A$ or the spectrum $\mathcal{E}_j$ of the modular Hamiltonian $H_A$ 
\be
\lambda_j\equiv e^{-\mathcal{E}_j},\qquad \sum_j \lambda_j=1,\label{ModSpecLE}
\ee
and, by analogy with thermal states, we can define the modular partition function at inverse temperature $\beta=n$ as
\be
\tilde{Z}(n)=\text{Tr}(\rho^n_A)=\sum_j e^{-n\mathcal{E}_j}.
\ee 
Conventionally, we normalise $\text{Tr}(\rho_A)=\tilde{Z}(1)=1$.

Finally, we define the modular evolution of the initial state \eqref{SchmDC} as
\begin{equation}
\left|\Psi(s)\right>=e^{-is H_A\otimes 1_{A^c}}\left|\Psi_0\right>,
\end{equation}
where $s$ is the modular time. Note that we perform this evolution with $H_A\otimes 1_{A^c}$ and not with the total modular Hamiltonian $H_{mod}=H_A\otimes 1_{A^c}-1_{A}\otimes H_{A^c}$; indeed, $\ket{\Psi_0}$ is invariant under the evolution with $H_{mod}$ \cite{Haag}. By analogy with the TFD state (evolution with $H_L$ vs $H_L-H_R$), this leads to a non-trivial evolution of the state $\ket{\Psi_0}$. In the following, our goal will be to quantify the spread complexity of this state in various models and shed light on the Lanczos coefficients in this evolution.

For that we use the Lanczos algorithm to construct an orthonormal basis $\ket{K_n}$ and expand our state as in \eqref{expand}, where the expansion coefficients $\psi_n(s)$ satisfy \eqref{SEq} with Lanczos coefficients $a_n$ and $b_n$ encoded in the modular return amplitude
\begin{equation}
S(s)\equiv \left<\Psi(s)|\Psi_0\right>=\sum_j \lambda_j^{1-is}=\tilde{Z}(1-is).
\end{equation}
This object is closely related to the Renyi entropies of the reduced density matrix $\rho_A$ defined for integer $n$ as
\be
S^{(n)}_A=\frac{1}{1-n}\log(\Tr\rho^n_A),
\ee
and we have the relation to the analytically continued Renyi with replica index $n=1-is$
\be
S(s)=\exp\left(is\,S^{(1-is)}_A\right).
\ee
We conclude that the Lanczos procedure, based on the moments of $S(s)$, will involve interesting combinations of quantum information measures. Indeed, already from \eqref{abmus}, we can see that for the modular Hamiltonian $H_A$, the moments $a_0$ and $b^2_1$ will be simply the von Neumann entropy $S_A$ and the capacity of entanglement $C_E$ \cite{Yao:2010woi,DeBoer:2018kvc,Kawabata:2021vyo,Okuyama:2021ylc,Arias:2023kni,Nandy:2021hmk} respectively. At the conceptual level, since spread complexity is a functional of the survival amplitude, and this is a functional of the entanglement spectrum, we conclude that, while entanglement is not enough (it is just $a_0$), entanglement spectrum is enough. Going in the reverse direction, since the Renyi entropies can be found from the modular survival amplitude, and this is a functional of the modular Lanczos spectrum, we also conclude that Lanczos spectrum is enough. This construction then provides a solid bridge between entanglement and complexity, as we further develop below.
\section*{Examples}
It is useful to consider a few simple, analytical examples. Let us start from a qubit state $\ket{\Psi_0}=\sqrt{p}\ket{00}+\sqrt{1-p}\ket{11}$ where $A$ and $A^c$ are the first and second spins respectively and $p\in[0,1]$. Tracing out the second Hilbert space we obtain the return amplitude
\be
S(s)=\Tr(\rho^{(1-is)}_1)=p^{(1-is)}+(1-p)^{(1-is)},
\ee
with moments (see definition \eqref{Moments}) 
\be
\mu_k=(-i)^k\left(p\log^k(p)+(1-p)\log^k(1-p)\right).
\ee
From them we extract the non-vanishing Lanczos coefficients
\bea
a_0&=& -p\log(p)-(1-p)\log(1-p) =S_1,\nonumber\\
b^2_1&=&p(1-p)\left(\log(1-p)-\log(p)\right)^2=C_E(\rho_1),\nonumber\\
a_1&=&-p\log(1-p)-(1-p)\log(p),
\eea
and confirm the relation with entanglement entropy and capacity of entanglement. At present, we do not have a sharp quantum information interpretation for $a_1$ and we hope to return to this issue in the future. 

Next, we derive the two solutions of the Schrodinger equation \eqref{SEq} with these Lanczos coefficients and they are 
\bea
\psi_0(s)&=&p^{1+is}+(1-p)^{1+is}=S(s)^*,\nonumber\\
\psi_1(s)&=&\mp\sqrt{p(1-p)}((1-p)^{is}-p^{is}),
\eea
with $\mp$ corresponding to $\pm$ in $b_1$ (recall that $b_n$'s \eqref{LanczCoeffDeff} are always positive so this sign depends on the difference between $\log(1-p)$ and $\log(p)$). By construction, the coefficient $\psi_0(s)$ and $S(s)$ are related by the simple complex conjugation.  Finally, the modular spread complexity \eqref{SpreadC} is given by
\be
\mathcal{C}(s)=4p(1-p)\sin^2\left(\frac{s}{2}\log\frac{1-p}{p}\right).
\ee
In this simple example with Krylov space dimension equal to $2$, we have a relation between the modular spread complexity and modular spectral form factor: $\mathcal{C}(s)=|\psi_1(s)|^2=1-|\psi_0(s)|^2$, which is not true in general for long times \cite{Erdmenger:2023shk}. Clearly, the complexity growth is determined by the value of $p$. In particular, for maximally entangled state $p=1/2$, the $b_1$ as well as $C(s)$ vanish (see Appendix B for another example). More generally for flat entanglement spectrum we have
\be
\Tr(\rho^n_A)=\text{dim}(\mathcal{H}_A)^{1-n},
\ee
and we only get non-trivial $a_0=S_A$ and all $b_n=0$. This is a physically sensible result. In this context we only need one number to understand the structure of the state. In the thermodynamic limit of physical systems, such as those appearing in quantum gravity, it might seem  that we have flat entanglement spectrum at micro-canonical sectors. This is only an artefact of the thermodynamic limit. In reality the spectrum is chaotic, and the eigenvalues, although close to the average flat value, show no degeneracy and resemble the spectrum of a random matrix. In this scenario the Lanczos spectrum is completely different, as we discuss below.

Another simple example consists of two coupled harmonic oscillators \cite{Srednicki:1993im}. After tracing one of them, we get the entanglement spectrum and modular partition function
\be
\lambda_k=(1-\xi)\xi^k,\qquad\tilde{Z}(n)=\frac{(1-\xi)^n}{1-\xi^n},
\ee
where $0<\xi<1$ is related to the details of the coupling between the oscillators. Following the above procedure, we can  derive a general form for the Lanczos coefficients
\bea
a_n&=&-n\frac{1+\xi}{1-\xi}\log(\xi)-\log(1-\xi)-\frac{\xi}{1-\xi}\log(\xi),\nonumber\\
b_n&=&n\frac{\sqrt{\xi}}{1-\xi}\log(1/\xi),
\eea
where again $a_0=S_1$ and $b^2_1=C_E(\rho_1)$. Observe that these Lanczos coefficients are governed by the SL(2,R) symmetry algebra and our modular evolution of the state can be mapped to a coherent state of this Lie group. This allows us to recycle the derivations in \cite{Caputa:2021sib} and derive the modular spread complexity
\be
\mathcal{C}(s)=\frac{4\xi}{(1-\xi)^2}\sin^2\left(\frac{s}{2}\log(\xi)\right).
\ee
The entanglement spectrum is equivalent to the thermal spectrum of a single oscillator, i.e. writing $\xi=\exp(-\beta \omega)$ we have $\lambda_k=e^{-\beta E_k}/Z(\beta)$ with $E_k$ being the energy of a single harmonic oscillator with frequency $\omega$. Even though we have an infinite dimensional Krylov basis, modular spread complexity oscillates. However, we can formally send $\omega\to i \tilde{\omega}$ (complex $\log(\xi)$) and observe exponential growth of the modular spread complexity.

Finally, we consider 2d CFT where the trace of the reduced density matrix of a single interval $A=[u,v]$ can be computed using the replica trick as a correlator of twist operators inserted at the end-points of $A$ \cite{Calabrese:2004eu}
\be
\tilde{Z}(n)=\langle\sigma_n(u)\tilde{\sigma}_n(v)\rangle=\exp(-\left(n-1/n\right)W),\label{PF2dCFT}
\ee
where $W$ contains the CFT central charge $c$ and details of the interval as well as geometry of the underlying CFT and is directly related to entanglement entropy $S_A=2W$ (e.g. $W=\frac{c}{6}\log((u-v)/\epsilon)$ for the vacuum in a line). For our discussion, we neglected an overall non-universal constant in \eqref{PF2dCFT}. However, it is crucial that we keep the cut-off $\epsilon$ small, but finite. The analytic continuation gives the modular partition function
\be
\tilde{Z}(1-is)=\exp\left(-\frac{s^2W}{s^2+1}+i\frac{s(s^2+2)W}{s^2+1}\right),\label{ACPF2dCFT}
\ee
therefore, the corresponding modular spectral form factor $|\tilde{Z}(1-is)|^2$ decays to a plateau with value $\exp(-S_A)$. 
By expanding Lanczos coefficients for large $W$ (or large central charge $c$, see Appendix C), we can show that spread complexity grows quadratically for initial modular time, proportionally to the entanglement entropy $S_A$
\be
\mathcal{C}(s)\sim S_A\, s^2.
\ee
For later times, at finite cut-off $\epsilon$, we also expect a period of linear growth and saturation to a plateau (analogous to the spectral form factor \eqref{ACPF2dCFT}).  Verifying this expectation numerically would be interesting and we leave it for future work. Next we move to more general qualitative arguments in the context of random matrix theory.
\section{Random Modular Hamiltonians}
Further relations between entanglement entropy and entanglement spectrum on one hand, and spread complexity and the Lanczos spectrum on the other, arise by considering the example of random pure states. Given a pure state and a bipartition of the system into $A$ and $A^c$, a putative ensemble of pure states (defining the particular notion of random state) naturally defines an ensemble of modular Hamiltonians $H_A$. This ensemble defines a particular notion of random modular Hamiltonian.

The analysis of the Lanczos approach for random matrices was recently developed in \cite{Balasubramanian:2022dnj,usrandom2}. The application of these constructions to modular evolution goes as follows. We first notice that, in the context of random states, the Lanczos coefficients of a reduced subsystem are random parameters, and the first goal is to compute their statistics. This can be accomplished with two assumptions. First we need take the thermodynamic limit, where the dimension $N$ of the subsystem $A$ goes to infinity. Without loss of generality we assume that this dimension is smaller than the dimension of $A^c$. In this limit the average values reliably inform us of the typical values associated with individual instances of the random modular Hamiltonian. Second we need to choose as initial state the vector $(1,0,\cdots, 0)$. The reason is that for this state we know how to compute the Jacobian of the transformation between the original form of the random modular Hamiltonian and the tridiagonal form. It is given by \cite{Balasubramanian:2022dnj}
\be
J=\prod\limits_{n=1}^{N-1} b_n^{(N-n)\beta-1}\;,
\ee
where $\beta$ is the Dyson index of the ensemble of random matrices. This Jacobian should be thought as the analogue of the Vandermonde determinant for the change of variables that takes us to the diagonal form of the matrix. Equivalently, if the ensemble is invariant under a certain group of unitaries, we are free to take any initial state that follows from the previous one by applying a unitary belonging to such a group.

In the thermodynamic limit, see \cite{PhysRevD.47.1640,doi:10.1063/1.533010,Balasubramanian:2022dnj,usrandom2}, it becomes natural to label the Lanczos coefficients in terms of $x\equiv n/N$, namely as $a(x)\equiv a_{n=xN}$ and $b(x)\equiv b_{n=xN}$. The reason is that in this limit, on average over the ensemble, the Lanczos coefficients become continuous functions in the interval $x\in [0,1]$. We can now obtain the relation between these functions and the modular spectrum. We cut the Krylov chain into shorter segments of a given length $L$, such that $L\rightarrow\infty$ and $L/N\rightarrow 0$ in the thermodynamic limit. This is a block approximation of the Hamiltonian whose density of states is the sum of the densities of each block. Given the continuity assumption, $a_n$ and $b_n$ can be taken as constants in each block, equal to $a(x)$ and $b(x)$. 

The different Hamiltonian blocks are then Toeplitz matrices of size $L$, with diagonal elements given by certain $a$ and off-diagonal elements given by certain $b$. These matrices have eigenvalues $E_k=2\,b\,\cos (k\pi/(L+1)) +a $, with $k=1,\cdots, L$, and their density of states read
\be \label{rhoseg}
\rho_{a,b}(E)=\frac{1/L}{|dE_k/dk|}= \frac{H(4\,b^2-(E-a)^2)}{\pi\,\sqrt{4\,b^2-(E-a)^2}}\;.
\ee
Here $H(x)$ is the Heaviside step function and we normalized the density of states by dividing by $L$. The total (normalized) density of states is the sum over all blocks. In the thermodynamic or continuum limit this becomes \cite{Balasubramanian:2022dnj}
\be \label{intdl}
\rho(E) =  \int_0^1 dx\, \frac{H(4\,b(x)^2-(E-a(x))^2)}{\pi\, \sqrt{4\,b(x)^2-(E-a(x))^2}}\;.
\ee
This formula relates the average Lanczos coefficients to the modular spectrum, in particular to the modular density of states, where we remind that $\lambda=e^{-E}$ (see \eqref{ModSpecLE}). Deviations from this formula were also found in \cite{Balasubramanian:2022dnj}, further providing a relation between the average Lanczos coefficients and the potential defining the ensemble of random matrices.

Generically, in chaotic systems the wavefunction in the Krylov basis \eqref{expand} reaches a stationary regime. In this regime the probabilities fluctuate around a mean value $\bar{p}(x)$. For special initial states we might have $\bar{p}(x)=1$, namely constant in $x$, but this is not the generic situation as can be established numerically in simple scenarios \cite{Balasubramanian:2022tpr,Balasubramanian:2022dnj}. It is thus natural to inquire for the form of the stationary distribution $\bar{p}(x)$. Indeed, in terms of the distribution of energies of the initial state
\be
\vert\,\langle\psi\vert E\rangle\,\vert^2 \equiv P(E)\;,
\ee
this is derived in \cite{usrandom2} as follows. Assume $P(E)$ is a continuous function of the energy. For the modular state evolution that we are considering, this implies a continuous entanglement spectrum with small fluctuations around the average. Using (\ref{rhoseg}), the number of states in the interval between $x$ and $x+dx$ and in the interval between $E$ and $E+dE$ is
\be \label{rhop}
\tilde{\rho}(x,E)\,dx\,
dE=  \frac{N\,dx\, dE}{\pi \,\sqrt{4b(x)^2-(E-a(x))^2}}.
\ee
The long-time average probability distribution in the Krylov basis is just the the convolution of this density with the distribution of energies of the initial state (which is conserved in time). This reads
\be
\bar{p}(x) = \int dE \,P(E)\,\tilde{\rho}(x,E)\,.
\ee
For the modular evolution of states the initial state was \eqref{SchmDC}. The distribution of energies in the initial state is then $P(E)=\lambda=e^{-E}$, and we arrive at
\be\label{pstfd} 
\bar{p}(x)=  \,N \,I_0(2 b(x))e^{- a(x)}\;.
\ee
The plateau of the modular spread complexity and the Shannon entropy $H_\text{Shannon}$ in the Krylov basis (dubbed K-entropy in \cite{Barbon:2019wsy}) follow from this probability distribution $\bar{p} (x)$. This is an explicit function once we have derived the Lanczos spectrum from the modular density of states using \eqref{intdl}. It turns out that the result for the Shannon entropy is quite insensitive to the specific ensemble of random modular Hamiltonian, i.e on the specific Lanczos coefficients $a(x)$ and $b(x)$. Indeed
\be
    H_\text{Shannon}= -\sum_n \bar{p}_n\ln \bar{p}_n\approx \ln N,
\ee
up to subleading corrections in the thermodynamic limit. This means that the dimension of the Hilbert space explored by the random modular evolution is the same as the number of non-zero eigenvalues in the reduced density matrix, counted by its leading density of states. Notice that this same result applies for the complementary subsystem $A^c$. Although we assumed the dimension of $A$ was smaller than that of $A^c$, the modular Hamiltonian and modular spectrum are the same up to zeros. In particular the number of non-zero eigenvalues is the same, and the saturation will happen at $\log N$ as well for $A^c$, where we remind that $N$ is the dimension of the smaller subsystem $A$.

Finally, we can turn things around. Starting from the Lanczos coefficientsf $a(x)$ and $b(x)$, we can find the stationary distribution of the modular spread complexity $\bar{p}(x)$ and from there we can obtain the initial probability distribution in the energy basis as
\be
P(E) = \int dx \,\bar{p}(x)\,\tilde{\rho}(x,E)\,.
\ee

We are led to the following conclusions. The first is that we could use these results in the context of the Page curve \cite{Page:1993df} (recall also that the relevance of the capacity of entanglement, that is our Lanczos coefficient $b_1$, to the Page curve was already discussed in \cite{Kawabata:2021vyo,Okuyama:2021ylc}). In this scenario, for random states drawn from the Haar measure the modular density of states is known and of compact support \cite{Page:1993df}. Although \eqref{intdl} cannot be solved in closed form in this case, the Lanczos $b(x)$ coefficients decay to zero as they should and the modular spread complexity follows the regimes described in \cite{Balasubramanian:2022tpr}. In particular, the spread complexity will saturate at a value controlled by the dimension of the smallest subsystem. For example, the entropy will be precisely $\log N$ in the leading approximation, where $N$ is such dimension. The plateau of modular spread complexity then draws a complexity Page curve in the same way as the entanglement entropy.

The second conclusion concerns the slogan ``entanglement is not enough'' \cite{Susskind:2014moa}. This was put forward to motivate the introduction of the notion complexity in quantum gravity. The present construction transparently shows why this is true when for the word ``entanglement'' we more precisely understand entanglement entropy itself. The reason is that entanglement entropy is the first entry of the Lanczos spectrum. But one needs the full spectrum of Lanczos coefficients to predict the long time dynamics of the wavefunction of the system. Spread complexity, which serves to characterize these dynamics, is also a functional of the whole spectrum. Clearly then, entanglement entropy is not enough. It is however not true if we slightly, but insightfully, modify the slogan so that it refers to the entanglement spectrum. As we have derived, there is a precise relation (one follows from the other and vice-versa), between the entanglement or modular spectrum, the modular Lanczos coefficients, the modular survival amplitude and the modular spread complexity. In this precise sense, we reach again the conclusion that the entanglement spectrum seems to be enough in the context of quantum gravity. The Lanczos modular spectrum and associated survival amplitude and modular complexity are enough as well.
\section*{Modular Growth and Evolution of Primary Operators}
In this final section we discuss the operator growth and spread complexity of operators under the modular flow with the total modular Hamiltonian $H_{mod}=H_A\otimes 1_{A^c}-1_{A}\otimes H_{A^c}$ ($A^c$ being the complement of $A$). Namely, we consider the following modular evolution \cite{Haag,Takesaki:1970aki,Borchers:2000pv}
\be
\mathcal{O}(s)=e^{iH_{mod}s}\mathcal{O}(0)e^{-iH_{mod}s}\equiv e^{i\mathcal{L}_{mod}s}\mathcal{O}(0),\label{ModEvOp}
\ee
where the modular Liouvillian (super-operator) acts on operators by taking the commutator $\mathcal{L}_{mod}\equiv [H_{mod},\cdot]$. This modular flow of operators has been a central topic in a variety of recent works in QFT and holography \cite{Hislop:1981uh,Casini:2009vk,Jafferis:2014lza,Jafferis:2015del,Faulkner:2017vdd,DeBoer:2019kdj,Mintchev:2022fcp,Lashkari:2018nsl,Leutheusser:2021frk,Casini:2011kv,Wong:2013gua,Cardy:2016fqc,Longo:2009mn,Hollands:2019hje} but, to our knowledge, its complexity remains relatively unexplored. 

To make progress, for simplicity, we first consider $H_{mod}$ for static, universal examples where $A$ is a single interval in the vacuum of a 2d CFT defined either on the line or on the circle, leaving more complicated cases to future works (see more in Appendix D). In the final part of this section we also mention the result for two disjoint intervals on the line for the free massless Dirac field in its ground state.

We start with the modular evolution of the highest weight state $\ket{h}$ (eigenstate of the CFT Hamiltonian i.e., $\ket{h}\equiv \lim_{z\to 0}\mathcal{O}(z)\ket{0}$ in radial quantisation of the Euclidean formalism) with the total modular Hamiltonian of an interval $A$ in 2d CFT 
\be
\ket{\psi(s)}=e^{-isH_{mod}}\ket{h}.\label{CHMod}
\ee
The total modular Hamiltonian is a well-defined operator in the continuum and, in 2d CFTs, it can be written as a linear combination of the  SL(2,R) generators (see e.g. \cite{Kabat:2017mun,Czech:2019vih})
\be
H_{mod}=\sigma_{-1}L_{-1}+\sigma_0 L_0+\sigma_1 L_1+a.c.,\label{ModHamSL2R}
\ee 
where the anti-chiral (a.c.) part is similarly expressed in terms of global $\bar{L}$'s (for simplicity, we will focus on the chiral part) and the coefficients depend on the CFT  and interval geometry (see Appendix D). For this reason \eqref{CHMod} is simply a coherent state and falls into the Lie-algebra symmetry examples considered in \cite{Caputa:2021sib,Balasubramanian:2022tpr} where, using the Baker–Campbell–Hausdorff formula, the spread complexity can be evaluated as a simple function of general $\sigma_i$'s (see \eqref{GenComplSL2R} in Appendix D). Before we write it down, note that we may think about this state simply in the context of spread complexity of states \cite{Balasubramanian:2022tpr} or as a state representing operator growth \cite{Parker:2018yvk} with a particular choice of the inner product that corresponds to the return amplitude
\be
S(s)=\langle h|e^{isH_{mod}}\ket{h}.\label{RetAmplGlobal2dh}
\ee
By using the procedures discussed in \cite{Caputa:2021sib,Balasubramanian:2022tpr}, we find that Lanczos coefficients from \eqref{RetAmplGlobal2dh} have the SL(2,R) form: $a_n=\gamma(n+\Delta)$ and $b_n=\alpha\sqrt{n(n+2\Delta-1)}$  with $\alpha=\sqrt{\sigma_1\sigma_{-1}}$ and $\gamma=\sigma_0$. Interestingly, in all the examples where \eqref{ModHamSL2R} holds, the coefficients satisfy $\sigma_1\sigma_{-1}-\sigma^2_0/4=\pi^2$ and this combination is directly linked to the Lyapunov exponent defined from the Krylov complexity \cite{Parker:2018yvk}. For example, for a single interval $A=[a,b]$ in 2d CFT on a circle of size $L$ we obtain
\be
\mathcal{C}(s)\simeq\frac{2h}{\sin^2\left(\frac{\pi(b-a)}{L}\right)}\sinh^2(\pi s).\label{ComPlGlobCH}
\ee
Clearly, at late modular time $s\gg 1$, the spread complexity grows exponentially with Lyapunov exponent $\lambda^{mod}_L=2\pi$. We will see below that this is in fact a universal behaviour also for local operator growth. The size of the entangling interval $b-a$ governs the scrambling time at late time (see also below). We should also point that, this result that uses $\sigma_i$'s from  \cite{Kabat:2017mun,Czech:2019vih} in the general formula \eqref{GenComplSL2R}, does not seem to have a well-defined (naive) limit of $L\to\infty$. As already pointed out in \cite{Caputa:2021sib,Caputa:2021ori}, the spread complexity of coherent states can be written as an expectation value of $L_0$. When passing from the cylinder to the plane, the derivative of the exponential map will bring the appropriate factor of $L$ that cures this (that is why we used $\simeq$).

Next, for 2d CFTs, we consider modular Hamiltonian evolution of states locally excited by a primary operator $\mathcal{O}(l)$ of conformal dimension $h$ placed inside the interval $A=[a,b]$, i.e., $\l\in A$. This state is defined as
\be
\ket{\psi(s)}=\mathcal{N}e^{-iH_{mod} s}e^{-\epsilon H}\mathcal{O}(l)\ket{0},\label{OurStateMH}
\ee
where $\ket{0}$ is the ground state of the entire system bipartite as $A\cup A^c$ and $H_{mod}$ is the total modular Hamiltonian associated with $A$ in this state. We remark that, again for the sake of simplicity, we only consider a chiral part of the 2d CFT.
Note that the local operator is first smeared with the CFT Hamiltonian by an amount $\epsilon$ in Euclidean time such that the energy of the excitation is finite $E_\mathcal{O}\sim h/\epsilon$ and factor $\mathcal{N}$ is the normalisation of this initial state with the operator. The standard Hamiltonian evolution of these states has been extensively studied in the past \cite{Nozaki:2014hna,He:2014mwa,Caputa:2014vaa,Caputa:2014eta} (see for corresponding spread complexity in Appendix D) but here we will be interested in the modular evolution instead. \\
Before we proceed, it is important to point that, since the operators $\mathcal{O}(l)$ are inserted in $A$, the actions of $H_{mod}$ and $H_A\otimes 1_{A^c}$ on them are identical. Hence, our discussion in the following also holds for \eqref{OurStateMH} with $H_{mod}$ replaced by $H_{A}$ and we will use them interchangeably in our formulas. In particular, the modular correlators that will be used in our return amplitudes (see below) are identical for these two modular evolutions.

Let us then recall a few basic facts about $H_A$. In the chiral 2d CFTs and for some particular states and bipartitions (e.g. when the CFT is defined either on the line or on the circle and is in its ground state), the modular Hamiltonian can be written as
\be
H_A=2\pi\int^b_a\beta_0(u)T(u)du,\qquad \beta_0(u)=\frac{1}{w'(u)},\label{ModHamMT}
\ee
where $T(u)$ is the chiral component of the 2d CFT energy-momentum tensor and the weight function $\beta_0(u)$ (often called local inverse temperature) encodes the dependence on the state and of the bipartition for the specific cases we are considering. For instance, for the ground state of a CFT on the line or on a circle of length $L$, we have respectively
\be
\label{w-explicit}
w(u) = \log\left(\frac{u-a}{b-u}\right),
\quad
w(u) = \log\left(\frac{\sin[\pi(u-a)/L]}{\sin[\pi(b-u)/L]}\right).
\ee
The modular evolution generated by (\ref{ModHamMT}) for a primary operator $\mathcal{O}$ of conformal dimension $h$ is
\be
\mathcal{O}(s,u)\equiv e^{is H_A}\mathcal{O}(u)e^{-is H_A},\label{ModEv2dCFT}
\ee
and it can be written as \cite{Hislop:1981uh, Casini:2011kv, Casini:2009vk}
\be
\label{mod-evo-O-explicit}
\mathcal{O}(s,u) = \left( \frac{\beta_0(\xi(s,u))}{\beta_0(u)} \right)^h \mathcal{O}\big(\xi(s,u)\big),
\ee
where $\mathcal{O}(u)$ is the initial configuration of the field at $s=0$ and $\xi(s,u) $ satisfies the following differential equation
\be
\label{pde-xi}
\partial_s \xi(s,u) = 2\pi \beta_0(x) \, \partial_u \xi(s,u),\qquad \xi(0,u) = u.
\ee
The solution of this equation reads
\be
\xi(s,u) \equiv w^{-1} \big( w(u) +2\pi s  \big),
\ee
in terms of $w(u)$ defined in (\ref{ModHamMT}) and its inverse function. \\
Then, the modular evolution (\ref{mod-evo-O-explicit}) can be expanded in powers of $s$ as follows
\be
\mathcal{O}(s,u) =
\sum_{n=0}^{\infty} 
\frac{(2\pi \, s)^n}{n!}\,
\widetilde{\mathcal{O}}_n(u).
\ee
By  employing (\ref{pde-xi}), the first three (non-trivial) operators in this expansion are
\bea
\label{O-tilde-123}
\widetilde{\mathcal{O}}_1(u) &=&\beta_0(u) \, \mathcal{O}'(u) 
+ h\, \beta'_0(u) \,  \mathcal{O}(u), 
\nonumber
\\
\rule{0pt}{.5cm}
\widetilde{\mathcal{O}}_2(u) &=&
\beta_0(u)^2 \,\mathcal{O}''(u) 
+ (2h+1) \beta_0(u)   \beta'_0(u) \,  \mathcal{O}'(u) \nonumber\\
&& 
+ \,h \big[ h \,\beta'_0(u)^2 + \beta_0(u) \beta''_0(u) \big] \mathcal{O}(u),
\nonumber\\
\rule{0pt}{.5cm}
\widetilde{\mathcal{O}}_3(u) 
&=&
\beta_0(u)^3 \,\mathcal{O}'''(u) 
+3(h+1) \beta_0(u)^2 \beta'_0(u) \,\mathcal{O}''(u) 
\nonumber \\
& &
+\,\beta_0(u) \big[ (3h^2 + 3h +1) \beta'_0(u)^2 \nonumber \\
& & \hspace{1.5cm}
+ (3h+1)\beta_0(u) \beta''_0(u) \big] \,\mathcal{O}'(u) 
\nonumber\\
& &
+\,h \big[ h^2 \beta'_0(u)^3 + (3h+1)\beta_0(u) \beta'_0(u)  \beta''_0(u) 
\nonumber \\
& & \hspace{.8cm}
+ \beta_0(u)^2  \beta'''_0(u) \big] \mathcal{O}(u) .
\eea
It is straightforward to write also $\widetilde{\mathcal{O}}_n(u)$ with $n >3$, 
 but their expressions are rather complicated to be reported here. \\
 Clearly, the growth of the operator (in operator space) due to the modular evolution \eqref{ModEv2dCFT} is determined also by the weight function $\beta_0(u)$ occurring in the modular Hamiltonian (\ref{ModHamMT}) and its non-trivial derivatives provide additional contributions of the initial field configuration $\mathcal{O}(u)$ into $\widetilde{\mathcal{O}}_n(u)$. Indeed, setting $\beta_0(u) = \textrm{const}$ in (\ref{O-tilde-123}) simplifies the expressions in a considerable way. On the other hand, the actual operator size and Krylov complexity \cite{Parker:2018yvk} is usually computed based on the return amplitude that, after an appropriate choice of the inner-product, may become a two-point correlator (computable with \eqref{mod-evo-O-explicit}). Below, we will add to these intuitions by computing the spread complexity of \eqref{OurStateMH} and find that it indeed depends on the local temperature $\beta_0(u)$.

Now, let us get back to the computation of the modular spread complexity of \eqref{OurStateMH}. The crucial ingredient is again the return amplitude that can be written as a special modular  two-point correlator
\be
S(s)=\frac{\langle\mathcal{O}^\dagger(0,u_1)\mathcal{O}(s,u_2)\rangle}{\langle\mathcal{O}^\dagger(0,u_1)\mathcal{O}(0,u_2)\rangle},
\ee
which satisfies $S(0)=1$ by construction. The insertion points of the operators in the initial state are (see Appendix D)
 \be
 u_1=l+i\epsilon,\qquad u_2=l-i\epsilon. \label{uis}
 \ee
 The two-point correlators of the operators after modular flow can be found e.g. in \cite{Hislop:1981uh,Casini:2009vk,Longo:2009mn,Hollands:2019hje}. Their general form is
\be
\frac{\langle\mathcal{O}(s_1,u_1)\mathcal{O}(s_2,u_2)\rangle}{\langle\mathcal{O}(0,u_1)\mathcal{O}(0,u_2)\rangle}=\left(\frac{e^{w(u_1)}-e^{w(u_2)}}{e^{w(u_1)+\pi s_{12}}-e^{w(u_2)-\pi s_{12}}}\right)^{2h},\label{Mod2ptfunMT}
\ee
where $s_{12} \equiv s_1-s_2$ and $w(u)$ is defined in \eqref{ModHamMT}. The modular correlator \eqref{Mod2ptfunMT} satisfies the KMS condition with inverse temperature $\beta_{KMS}=1$.  Using \eqref{Mod2ptfunMT}, we can write our return amplitude with general $w(u)$ as
\be
S(s)=\left(\frac{e^{-\pi s}(1-B)}{e^{-2\pi s}-B}\right)^{2h},\qquad B=e^{w(u_2)-w(u_1)},\label{RetAmplUnivMT}
\ee
where $B$, via $w(u)$, depends on the details of the bipartition. This return amplitude again falls into the SL(2,R) symmetry class and we can derive universal Lanczos coefficients $a_n$ and $b_n$ for arbitrary $B$ (or $w(u)$) and compute the modular spread complexity. To derive correct Lanczos coefficients, it is important to perform this computation for general $B$ and take small $\epsilon$ in \eqref{uis} only at the end (instead of first expansing $S(s)$ in $\epsilon$ and then trying to derive moments). The Lanczos coefficients are 
\bea
a_n&=&\frac{2\pi i (B+1)}{B-1}(n+h),\nonumber\\
b_n&=&\frac{2\pi\sqrt{B}}{\sqrt{-(B-1)^2}}\sqrt{n(n+2h-1)},
\eea
where the factor of $i$ and the signs are chosen such that they are real for our physical insertion points \eqref{uis}. Finally, the spread complexity for finite $\epsilon$ can be written compactly for arbitrary $w(u)$ as
\be
\mathcal{C}(s)=\frac{2h}{\sin^2\left(\frac{i(w(u_1)-w(u_2))}{2}\right)}\sinh^2(\pi s).
\ee
Comparing with \eqref{ComPlGlobCH}, we see that the modular evolution with $\sinh^2(\pi s)$ is universally the same but the pre-factor in the present case involves the details of the insertion of the local operator $\mathcal{O}(l)$.\\
Interestingly, the small $\epsilon$ expansion leads to 
\be
\mathcal{C}(s)=2h\frac{\beta_0(l)^2}{\epsilon^2}\sinh^2(\pi s)+O(\epsilon^0),
\ee
which depends on $\beta_0(l)$, while the sub-leading orders contain also the derivatives of $\beta_0(l)$ (see \eqref{EpsExpwFT}).\\ 
We remark that the universal dependence on $s$ is consistent with the analyticity properties of the two-point function \eqref{Mod2ptfunMT} and, since the KMS inverse temperature is $\beta_{KMS} =1$ for the modular evolution,  it can be understood as $\sinh^2(\pi s/\beta_{KMS} )$. Moreover, at late modular time $s$, we find
\be
C(s)\sim e^{\lambda^{mod}_L\left(s-s_*\right)},
\ee
where the modular Lyapunov exponent $\lambda^{mod}_L$ and scrambling time  $s_*$ for the local operator are determined by the local temperature of the modular Hamiltonian respectively as
\be
\lambda^{mod}_L=2\pi,\qquad s_*=\frac{1}{\pi}\log\left(\sqrt{\frac{2}{h}}\frac{\epsilon}{\beta_0(l)}\right).
\ee
It is interesting to point that, since the spatial bipartition is symmetric w.r.t. the center of the interval, the coefficient of the modular spread complexity $\beta_0(l)^2$ (or the scrambling time) is maximal for $l$ in the middle of the entangling region $l=(a+b)/2$ whereas it is suppressed (vanishes) close to the boundary points of the entangling interval (see e.g. \eqref{BetaFSIZE}).
These are our main results in this section. Similarly to the bound of the Lyapunov exponent from Krylov complexity \cite{Parker:2018yvk,Gu:2021xaj} we conjecture that our modular exponent provides a bound on the modular chaos (see e.g. \cite{DeBoer:2019kdj}).

For more intervals, general modular Hamiltonians become more complicated and non-universal so the analysis is beyond the scope of this work. 
Nevertheless, for the free massless Dirac fermion in the vacuum, the modular Hamiltonian of disjoint intervals and the dicorresponding two-point modular correlators 
are known explicitly \cite{Casini:2009vk}. 
More precisely, we can consider local  fermion operator $\Psi(l)$ with $h=1/2$, in either of the two intervals $[a_1, b_1]\cup [a_2, b_2]$, evolved with the modular Hamiltonian of this union region. The two-point function of modular flow of $\Psi$ is known in this case \cite{Casini:2009vk, Longo:2009mn} (see also e.g. \cite{Hollands:2019hje}) 
and, somewhat surprisingly, it turns out that the corresponding return amplitude can again be written as \eqref{RetAmplUnivMT} with
\be
w(u)=\log\left(-\frac{(u-a_1)(u-a_2)}{(u-b_1)(u-b_2)}\right).\label{Mapsw2int}
\ee
This is sufficient to determine the modular spread complexity that, in the leading $\epsilon$, becomes 
\be
\mathcal{C}(s)=\frac{\beta^{loc}_0(l)^2}{\epsilon^2}\sinh^2(\pi s),\qquad \beta^{loc}_0(u)=\frac{1}{w'(u)},
\ee
in terms of \eqref{Mapsw2int}. In fact the local part of this modular Hamiltonian (that also contains a non-local piece) of these two disjoint intervals can again be written in the form \eqref{ModHamMT} with $\beta^{loc}_0(u)$; hence it governs the scrambling time. 
\section*{Discussion and Outlook}
In this work we have expanded Krylov complexity technology to the context of modular evolution. In particular, we have studied the relations between the entanglement spectrum, the Lanczos spectrum, and the notions of Krylov and spread complexity in various concrete examples. On one hand, this construction transparently shows why ``entanglement is not enough'' \cite{Susskind:2014moa}. In fact, from the complexity perspective of this story, entanglement entropy is just the first Lanczos coefficient, namely $a_0=S_E$. However, to understand the evolution of the wavefunction, and consequently of spread complexity, we need to know the full modular Lanczos spectrum. One can make an analogous statement about the TFD state, where the thermal entropy is related to the first Lanczos coefficient, but all the higher coefficients are also crucial to determine the evolution of complexity. On the other hand, the full Lanczos spectrum is obtained from the entanglement spectrum, providing concrete evidence that entanglement spectrum may be enough in certain scenarios.

From a different standpoint, from the Lanczos spectrum one can determine the entanglement spectrum up to degeneracies. In fact one can obtain all the moments of the modular Hamiltonian, and therefore the modular flow and all Renyi entropies. This way, the analysis of the modular Lanczos coefficients opens up a new window on the study of entanglement measures. As we show, the entanglement entropy is the first Lanczos coefficient while the capacity of entanglement is the second. It would be interesting if also the higher $n$ Lanczos coefficients contain similar information theoretic interpretations (perhaps along the lines of entanglement monotones \cite{Arias:2023duc}) and we leave this problem for the future investigation.

Then, we found that the modular growth of operators exhibits universal modular Lyapunov exponent $\lambda^{mod}_L=2\pi$, related to the $\beta_{KMS}$ and analyticity of the return amplitude, as well as the scrambling time sensitive to the local temperature of the CFT modular Hamiltonians. Going beyond our universal examples is certainly very important. For example, numerics for modular Hamiltonians in lattice models \cite{Peschel:2003rdm,Eisler-Peschel-09-review,Casini:2009sr}, would clarify the aforementioned relation between entanglement, complexity and modular chaos. 

In addition, we remark that in our analysis of spread complexity, by definition, we work with the standard, natural inner product in Hilbert space. For Krylov complexity of operators \cite{Parker:2018yvk}, the freedom of choosing a different inner product (e.g. Wightman) provides a different modular Krylov complexity. A systematic study and better understanding of sensitivity of modular complexity and modular chaos to these choices is an interesting open problem.

Another important direction concerns holography. In holographic theories we expect the relation \cite{Jafferis:2015del}
\be
H^{bdy}_{mod}=\frac{\hat{A}_{ext}}{4G_N}+\hat{S}_{\text{Wald-like}}+H^{bulk}_{mod}\;,
\ee
for the boundary modular Hamiltonian in terms of bulk quantities ($\hat{A}_{ext}$ computes the area of the Ryu-Takayanagi \cite{Ryu:2006bv} extremal surface $\mathcal{S}$ in the bulk and $\hat{S}_{\text{Wald-like}}$ can be expressed by expectation values of local operators on $\mathcal{S}$ \cite{Jafferis:2015del}). In the semiclassical limit, we also have that in the entanglement wedge of region $R_b$ the commutators $[H^{bdy}_{mod},\phi_R]$ and $[H^{bdy}_{bulk},\phi_R]$ are the same, for any local operator $\phi_R$ in $R_b$. Equivalently, in the low energy limit (in the code-subspace) we have
\be
\phi_R(s)=e^{is\mathcal{L}_{bdy}}\phi_R=e^{is\mathcal{L}_{bulk}}\phi_R,
\ee
and the Krylov subspace with modular Liouvillians $\mathcal{L}_{bdy}$ or $\mathcal{L}_{bulk}$ will be the same. It is then interesting to study the complexity of bulk reconstruction joining the results developed in \cite{Faulkner:2017vdd}, the analysis of the Lanczos approach for generalized free fields described in \cite{Magan:2020iac}, and the present techniques. This might naturally be extended to the complexity of extracting information from the black hole interior, following the islands construction \cite{Penington:2019kki}.

Finally, it would also be interesting to extend our discussion to the analysis of the black hole micro-states put forward in \cite{Balasubramanian:2022gmo,Balasubramanian:2022lnw}, which are insightful examples of the so-called PETP states \cite{Goel:2018ubv}. We hope to report on it in the near future.

\subsection*{Acknowledgements} 
We are grateful to Vijay Balasubramanian, Jan Boruch, Anatoly Dymarsky,  Nima Lashkari, Sinong Liu, Joan Simon, Qingyue Wu and Claire Zukowski for many conversations on Krylov and spread complexity, and useful comments on the present draft. We also wish to thank Roberto Auzzi, Shira Chapman, Aldo Cotrone, Dongsheng Ge, Francesco Gentile, Mihail Mintchev, Giuseppe Mussardo, Giuseppe Policastro, Domenico Seminara. PC and DP are supported by NAWA “Polish Returns 2019” PPN/PPO/2019/1/00010/U/0001 and NCN Sonata Bis 9 2019/34/E/ST2/00123 grants. The work of JM is supported by CONICET, Argentina. ET is grateful to the Henri Poincar\'e Institute (Paris) and to the CTP at MIT (Boston) for hospitality and financial support during part of this work.



\section{Appendix A: Lanczos coefficients and moments}\label{AppendixA}
Here we briefly explain how the Lanczos coefficients can be computed in a simple way (for relatively low $n$). By definition the moments of the return amplitude $S(t)$ (for real or modular times $t$ or $s$ respectively) are related to the expectation value of the evolving Hamiltonian 
\be
\mu_n=\langle \psi_0|(iH)^n|\psi_0\rangle.\label{MnHnRel}
\ee
Also, by construction, the initial state is the first state in the Krylov basis $|\psi_0\rangle=\ket{K_0}$ in which $H$ is tri-diagonal with Lanczos coefficients $a_n$ on the diagonal and off-diagonal $b_n$'s. The recursive algorithm is a version of a Markov process where $\mu_n$ for some fixed $n$ is expressed only in terms of lanczos coefficients $a_n$ and $b_n$ with labels only up to that fixed $n$. This way, we can find polynomial relations between $\mu_n$'s and Lanczos coefficients directly from \eqref{MnHnRel} by simply generating a tri-diagonal matrix $H$ of at least size $n$, taking its $n$-th power and extracting the $00$-element. This gives
\be
\mu_1=ia_0,\qquad \mu_2=-a^2_0-b^2_0,
\ee
and
\be
\mu_3=-i(a^3_0+2a_0b^2_1+a_1b^2_1),
\ee
and so on. Then we just solve these relations remembering that $b_n$ are the positive normalisations of the Krylov basis states. This gives 
\bea
a_0=-i\mu_1,\qquad b^2_1=\mu^2_1-\mu_2,
\eea
and
\be
a_1=i\frac{\mu^3_1-2\mu_1\mu_2+\mu_3}{\mu^2_1-\mu_2},
\ee
\be
b^2_2=\frac{\mu^3_2+\mu^2_3+\mu^2_1\mu_4-2\mu_1\mu_2\mu_4-\mu_2\mu_4}{(\mu^2_1-\mu_2)^2}.
\ee
For simple return amplitudes (e.g. fixed by dynamical Lie algebra symmetry) we can often guess a general form after first several steps of this procedure and verify that it holds for higher $n$'s. 

The polynomial relations above are just part of the algorithm, and hold for arbitrary $\mu_n$, $a_n$ and $b_n$ in the Lanczos algorithm. For a more detailed recursive derivation of the Lanczos coefficients in terms of the survival amplitude and moments, see \cite{Balasubramanian:2022tpr}.

\section{Appendix B: GHZ vs W-states}\label{AppendixB}
To gain more intuition for modular spread complexity in multi-partite setups, here we give one more example in quantum mechanics with tripartite entangled states of class GHZ and W. Let us start from slightly more general states parametrized as
\be
\ket{GHZ}_p=\sqrt{p}\ket{000}+\sqrt{1-p}\ket{111},
\ee
and
\be
\ket{W}_{p_1,p_2}=\sqrt{p_1}\ket{100}+\sqrt{p_2}\ket{010}+\sqrt{1-p_1-p_2}\ket{001}.
\ee
If we trace one of the spins in the first case, we end up with the reduced density matrix of the two spins with modular eigenvalues $\{p,1-p\}$ (studied in the main text) and spread complexity
\be
\mathcal{C}_p(s)=4p(1-p)\sin^2\left(\frac{s}{2}\log\frac{1-p}{p}\right).\label{MSCOMGHZW}
\ee
For the actual GHZ state with $p=1/2$, the complexity vanishes.

On the other hand, tracing out the first or second spin in the W-class states brings again the reduced density matrix for two spins with modular eigenvalues $\{p_i,1-p_i\}$, $i=1,2$ and modular spread complexity \eqref{MSCOMGHZW} with $p_i$. Moreover, integrating over the third spin gives modular spectrum $\{p_1+p_2,1-(p_1+p_2)\}$ and modular spread complexity
\be
\mathcal{C}_{p_1,p_2}(s)=4(p_1+p_2)(1-p_1-p_2)\sin^2\left(\frac{s}{2}\log\frac{1-p_1-p_2}{p_1+p_2}\right).
\ee
For the W-state with $p_1=p_2=1/3$ we get
\be
C(s)=\frac{8}{9}\sin^2\left(\frac{s}{2}\log(2)\right).
\ee
Clearly, the structure of entanglement is more susceptible to increase in complexity for the W-state that is not maximally entangled. It may be interesting to repeat this analysis more generally in the Hilbert space of 3 qubits for a state where we sum over all the basis vectors (probably numerically) with appropriate coefficients. 

\section{Appendix C: Lanczos coefficients for large W }\label{AppendixC}
In the main text, we have used the expansion of Lanczos coefficients coming from \eqref{ACPF2dCFT} for large values of $W$. This was done as follows. First we can compute several coefficients exactly and they read
\be
a_0=2W,\quad a_1=2W+3,\quad b^2_1=2W,\quad b^2_2=4W+3,
\ee
however, from $a_2$ they get more complicated:
\bea
a_2&=&\frac{8W^2+30W+15}{4W+3},\nonumber\\
b^2_3&=&6\frac{16W^3+48W^2+45W+12}{(4W+3)^2},
\eea
and so on. What we can do in practice is to take these exact solutions (up to some large say $n\sim 15$) and expand them for large $W$. From this we analytically find general answer for the first couple of orders
\bea
b_n&=& \sqrt{2W}\sqrt{n}\left[1+\frac{3(n-1)}{8W}-\frac{9(n-1)^2}{128W^2}+O(W^{-3})\right],\nonumber\\
a_n&=&2W\left[1+\frac{3n}{2W}-\frac{3n(n-1)}{16W^2}+O(W^{-3})\right].\label{anbnExp}
\eea
We can see that for initial $n\ll W$ we have constant $a_n\sim 2W$'s and $b_n\sim\sqrt{2W}\sqrt{n}$. We can then solve the Schrodinger equation \eqref{SEq} for $\psi_n(s)$ in this regime by first noting that, for constant $a_n=a$, we can simply substitute
\be
\psi_n(s)=e^{-ia s}\phi_n(s),
\ee
with $\phi_s(s)$ satisfying
\be
i\partial_s\phi_n(s)=b_n\phi_{n-1}(s)+b_{n+1}\phi_{n+1}(s).
\ee
More generally, for $b_n=\alpha\sqrt{n}$, the solution of this equation is simply the same as for the Heisenberg-Weyl algebra \cite{Caputa:2021sib}
\be
\phi_n(s)=\frac{(-i\alpha s)^n}{\sqrt{n!}}e^{-\frac{1}{2}\alpha^2s^2},
\ee
and the spread complexity grows quadratically in this initial regime with coefficient specified by the entanglement entropy $S_A=2W$
\be
\mathcal{C}(s)=\alpha^2 s^2=2W\,s^2=S_A\, s^2.
\ee
When $n\sim W$, this expansion breaks down. Beyond this regime a transition to linear growth of complexity followed by a saturation is expected because the modular spectral form factor saturates.  It would be very interesting to verify these two behaviours explicitly with numerics and we leave it as an important future problem.

\section{Appendix D: Local Operator Evolution}\label{AppendixD}
In this appendix we provide more details for the spread complexity of local operators under modular evolution. In order to gain some intuition and perspective on this computation, we first evaluate spread complexity of a state locally excited by a primary operator and evolved with the Hamiltonian of a 2d CFT. This setup has been extensively studied before \cite{Nozaki:2014hna,He:2014mwa,Caputa:2014vaa,Caputa:2014eta} as a milder version of a local quench. 
\subsection{Hamiltonian evolution}
The starting point is a quantum state locally excited by a primary operator of conformal dimensions $\Delta=h+\bar{h}$ inserted in position $l$. The density matrix can be written as
\be
\rho_0=\mathcal{N}e^{-\epsilon H}\mathcal{O}(l)\ket{0}\bra{0}\mathcal{O}^\dagger(l)e^{-\epsilon H}\equiv \ketbra{\psi_0},
\ee
where we regulate (smear) the operator in Euclidean time with cut-off $\epsilon$ that makes the energy of the excitation finite $(E_\mathcal{O}\sim\Delta/\epsilon)$. The normalisation is chosen such that Tr($\rho$)=1. In the previous studies, one was interested in the real time evolution of $\rho(t)=e^{-iHt}\rho_0 e^{iHt}$ and dynamics of entanglement or correlation functions in this protocol. Here we first focus on the spread complexity of the associated state
\be
\ket{\psi(t)}=e^{-iHt}\ket{\psi_0},
\ee
which can be computed from the return amplitude
\be
S(t)^*=\text{Tr}(\rho_0e^{-iHt})=\frac{\langle \mathcal{O}^\dagger(z_1,\bar{z}_1)\mathcal{O}(z_2(t),\bar{z}_2(t))\rangle}{\langle\mathcal{O}^\dagger(z_1,\bar{z}_1) \mathcal{O}(z_2(0),\bar{z}_2(0))\rangle},
\ee
where we used complex coordinates $(z,\bar{z})=(x+i\tau,x-i\tau)$ in which the insertion points are
\bea
&&z_1=l+i\epsilon,\quad \bar{z}_1=l-i\epsilon,\nonumber\\
&&z_2(t)=l-i(\epsilon+it),\quad \bar{z}_2(t)=l+i(\epsilon+it).
\eea
If we start from a two-point correlator in a CFT on a line 
\be
\langle\mathcal{O}^\dagger(z_1,\bar{z}_1) \mathcal{O}(z_2,\bar{z}_2)\rangle=z^{-2h}_{12}\bar{z}^{-2\bar{h}}_{12},
\ee
with $z_{ij}=z_i -z_j$, the return amplitude becomes
\be
S(t)=\left(1-\frac{it}{2\epsilon}\right)^{-2\Delta}.\label{retSInf}
\ee
Analogously, using the two-point function in a CFT on a circle of size $L$ 
\be
\langle\mathcal{O}^\dagger(z_1,\bar{z}_1) \mathcal{O}(z_2,\bar{z}_2)\rangle=\left(\frac{L}{\pi }\sin\left(\frac{\pi z_{12}}{L}\right)\right)^{-2h}\left(...\right)^{-2\bar{h}},
\ee
yields the return amplitude 
\be
S(t)=\left(\frac{\sinh\left(\frac{2\pi\epsilon}{L}\left(1-\frac{it}{2\epsilon}\right)\right)}{\sinh\left(\frac{2\pi\epsilon}{L}\right)}\right)^{-2\Delta}.\label{ReAmpL}
\ee
It oscillates in time and, it reduces to \eqref{retSInf} in the infinite $L$ limit. Since both correlators are translationally invariant, the dependence on the insertion point $l$ cancels. We could also introduce a UV cut-off $\epsilon_{UV}$ to the correlators but it would also cancel in the normalisation of the return amplitude. However, the return amplitude clearly depends on the operator regulator $\epsilon$. Finally, we can also use the two-point correlator at finite temperature that is formally obtained by taking $L=i\beta$. We will then evaluate spread complexity from \eqref{ReAmpL} and simply extract answers for the CFT on the line and at finite temperature by the above limit and the substitution. The last important remark is that, since we are interested in the evolution for all times, we should derive Lanczos coefficients and compute the spread complexity with finite $\epsilon$ and consider small $\epsilon$ only at the end of the computation. The other order of limits, taking first small $\epsilon$ expansion of $S(t)$ and then computing moments and Lanczos coefficients, is simply incorrect.

Following the algorithm \cite{Balasubramanian:2022tpr,LanczosVish} we can compute the moments and extract Lanczos coefficients analytically
\bea
a_n&=&\frac{2\pi\,(n+\Delta)}{L\tanh\left(\frac{2\pi\epsilon}{L}\right)},\nonumber\\
b_n&=&\frac{\pi}{L\sinh\left(\frac{2\pi\epsilon}{L}\right)}\sqrt{n(n+2\Delta-1)},\label{anbnLO}
\eea
and they correspond to the Lanczos coefficients governed by the SL(2,R) algebra labeled by the highes weight representation $\Delta$  that have a general form $a_n=\gamma(n+\Delta)$ and $b_n=\alpha\sqrt{n(n+2\Delta-1)}$ \cite{Caputa:2021sib}. In that case the spread complexity is a general function of $\alpha$ and $\gamma$ and becomes \cite{Balasubramanian:2022tpr}
\be
\mathcal{C}(t)=\frac{2\Delta}{1-\frac{\gamma^2}{4\alpha^2}}\sinh^2\left(t\sqrt{\alpha^2-\frac{\gamma^2}{4}}\right).\label{GenComplSL2R}
\ee
For our coefficients \eqref{anbnLO}, we then have
\be
\mathcal{C}(t)=2\Delta\frac{\sin^2\left(\frac{\pi t}{L}\right)}{\sinh^2\left(\frac{2\pi \epsilon}{L}\right)}.
\ee
Clearly it oscillates in time, with period $L$, and in the small $\epsilon$ limit it is proportional to the energy of the excitation. For large $L$ it reproduces the answer for the CFT on a line and grows quadratically with time
\be
\mathcal{C}(t)=\frac{\Delta }{2\epsilon^2}t^2.
\ee
This is also consistent with the limiting behaviour of $\alpha$ and $\gamma$ that satisfy $\gamma=2\alpha$ when $L\to\infty$. Lastly, continuing to $L\to i\beta$ gives exponentially growing spread complexity
\be
\mathcal{C}(t)=2\Delta\frac{\sinh^2\left(\frac{\pi t}{\beta}\right)}{\sin^2\left(\frac{2\pi \epsilon}{\beta}\right)}.
\ee
At late times, this complexity is characterized by the same the Lyapunov exponent $\lambda=2\pi/\beta$ as obtained from the Krylov complexity of the operator growth \cite{Parker:2018yvk}.
\subsection{Modular Hamiltonian evolution}
We can now move to the modular Hamiltonian evolution. The starting point will be the same as above i.e., locally excited state $\ket{\psi_0}$, but we consider the insertion $l$ of the local operator $\mathcal{O}$ to be inside an interval $A=[a,b]$ and perform the evolution with the modular Hamiltonian of $A$ of the initial state. For the vacuum excitations, the state of our interest will be
\be
\ket{\psi(s)}=\mathcal{N}e^{-iH_A s}e^{-\epsilon H}\mathcal{O}(l)\ket{0},
\ee
where we denote the modular time as $s$ and evolution is done with the modular Hamiltonian of $A$ 
\be
H^\pm_A=\pm 2\pi\int^b_a\beta_0(u_\pm)T_\pm(u_\pm)du_\pm,\qquad \beta_0(u)=\frac{1}{w'(u)},\label{ModHam1}
\ee
where $T_\pm$ are chiral and anti-chiral components of the energy-momentum tensor. For simplicity we will consider chiral operators and drop the $\pm$ notation. The kernel $\beta_0(u)$ stands for effective (inverse) temperature and e.g. for the CFT on a circle of size $L$  and in its ground state is given by
\be
\beta_0(u)=\frac{L}{\pi}\frac{\sin\frac{\pi(b-u)}{L}\sin\frac{\pi(u-a)}{L}}{\sin\frac{\pi(b-a)}{L}}.\label{BetaFSIZE}
\ee
Note that for the operator $\mathcal{O}(l)$ inside $A$ we could evolve with the total modular Hamiltonian of $A$ and its complement $A^c$ defined as $H_{mod}=H_A-H_{A^c}$ that has the same action on operators in $A$ but is a well-defined operator in continuum QFT. In 2d CFT, this total modular Hamiltonian for a single interval can be written in terms of global SL(2,R) generators (see e.g. \cite{Kabat:2017mun,Czech:2019vih})
\be
H_{mod}=\sigma_{-1}L_{-1}+\sigma_0L_0+\sigma_1L_1+a.c.,\label{TotModHam}
\ee
where the coefficients $\sigma_i$ depend on the end-points of the interval and the size (or temperature) and $a.c.$ stands for the anti-chiral part in terms of $\bar{L}_n$'s.  For example, the coefficients for a single interval $A=[a,b]$ in CFT on a circle of size $L$ are
\be
\sigma_0=-2\pi\cot\frac{\pi(b-a)}{L},\qquad \sigma_{\pm1}=\frac{2\pi\cot\frac{\pi(b-a)}{L}}{e^{\pm\frac{2\pi i }{L}b}+e^{\pm\frac{2\pi i }{L}a}}.
\ee
Generally, these types of SL(2,R) Hamiltonians describe inhomogeneous 2d CFTs and can be understood geometrically as different quantisation (than e.g. the usual radial used in CFTs).

The important ingredient is again the return amplitude
\be
S(s)=\bra{\psi_0}e^{iH_A s}\ket{\psi_0}=\frac{\langle\mathcal{O}^\dagger(0,u_1)\mathcal{O}(s,u_2)\rangle}{\langle\mathcal{O}^\dagger(0,u_1)\mathcal{O}(0,u_2)\rangle},
\ee
where $\mathcal{O}(s,u)\equiv e^{is H_A}\mathcal{O}(u)e^{-is H_A}$ is  the chiral flow of the operator $\mathcal{O}(u)$ and $u_1=l+i\epsilon$ and $u_2=l-i\epsilon$. The relevant two-point correlators of the operators after modular flow can be found e.g. in \cite{Mintchev:2022fcp}. Their general form is given in \eqref{Mod2ptfunMT} and they satisfy the KMS condition with periodicity $s+i$ so are analytic on a strip of size $\beta_{KMS}=1$. This way, we can write our return amplitude as
\be
S(s)=\left(\frac{e^{-\pi s}(1-B)}{e^{-2\pi s}-B}\right)^{2h},\qquad B=e^{w(u_2)-w(u_1)},\label{RetAmplUniv}
\ee
where $B$ depends on the details of the bipartition via $w(u)$. Interestingly, we can show that moments of this amplitude again correspond to the SL(2,R) Lanczos coefficients that are real, and can be written in terms of $B$ as
\bea
a_n&=&\frac{2\pi i (B+1)}{B-1}(n+h),\nonumber\\
b_n&=&\frac{2\pi\sqrt{B}}{\sqrt{-(B-1)^2}}\sqrt{n(n+2h-1)}.
\eea
Equivalently, one can obtain these coefficients also by applying the procedure described in \cite{Dymarsky:2021bjq} to the modular correlators (with the convention that our $b_n$ corresponds to $b_{n+1}$ in that work; for us $b_0=0$).\\
This allows us to read off $\alpha$ and $\gamma$ as before and check that they satisfy
\be
\alpha^2-\gamma^2/4=\pi^2,\quad 1-\frac{\gamma^2}{4\alpha^2}=\frac{-(B-1)^2}{4B}=\sin^2\left(\frac{iw_{12}}{2}\right),\label{2IdenT}
\ee
with $w_{12}=w(u_1)-w(u_2)$. For our insertion points $u_1=l+i\epsilon$ and $u_2=l-i\epsilon$, we then find 
\be
1-\frac{\gamma^2}{4\alpha^2}=w'(l)^2\epsilon^2+O(\epsilon^4).
\ee
The coefficient of the first term is given by the effective temperature evaluated at the operator's insertion point $w'(l)=1/\beta_0(l)$. Combining the two identities \eqref{2IdenT}, we get the universal result for the modular spread complexity of local operators for small $\epsilon$ quoted in the main text
\be
\mathcal{C}(s)=2h\frac{\beta_0(l)^2}{\epsilon^2}\sinh^2(\pi s).
\ee
For finite $\epsilon$ we simply get
\be
\mathcal{C}(s)=\frac{8h\,B}{-(B-1)^2}\sinh^2(\pi s)
=\frac{2h}{\sin^2\left(\frac{iw_{12}}{2}\right)}\sinh^2(\pi s).
\ee
We can expand it in $\epsilon$ as (suppressing $l$ dependance)
\be
\mathcal{C}(s)=\left[\frac{\beta_0^2}{\epsilon^2}+\frac{1+2\beta'^2_0-\beta_0\beta''_0}{3}+O(\epsilon^2)\right]2h\sinh^2(\pi s),\label{EpsExpwFT}
\ee
so we see that the sub-leading corrections also depend on derivatives of $\beta_0(l)$.\\
The finite $\epsilon$ expression can be also written explicitly in our examples, e.g. for finite size we have
\be
\frac{1}{\sin^2\left(\frac{iw_{12}}{2}\right)}=\frac{4\left|\sin\left(\frac{\pi (b-l+i\epsilon)}{L}\right)\right|^2\left|\sin\left(\frac{\pi (l-a+i\epsilon)}{L}\right)\right|^2}{\sin^2\left(\frac{\pi (b-a)}{L}\right)\sinh^2\left(\frac{2\pi \epsilon}{L}\right)}.
\ee
The finite temperature case is recovered by $L=i\beta$ and for the CFT on a line and in the vacuum we simply get
\be
\frac{1}{\sin^2\left(\frac{iw_{12}}{2}\right)}=\frac{((b-l)^2+\epsilon^2)((l-a)^2+\epsilon^2)}{(b-a)^2\epsilon^2}.
\ee

\end{document}